\documentclass{osa-article}

\journal{osajournal}

\articletype{Research Article}

\usepackage{units}

\begin{document}

\title{Non-equilibrium dynamics in the dual-wavelength operation of Vertical external-cavity surface-emitting lasers}

\author{I.~Kilen\authormark{1,*}, J.~Hader\authormark{1,2}, S.~W.~Koch \authormark{1,3}, and J.~V.~Moloney \authormark{1,2,4}}

\address{
\authormark{1}College of Optical Sciences, University of Arizona, 1630 East University Boulevard, Tucson, Arizona 85721, USA\\ \authormark{2}Nonlinear Control Strategies Inc., 7040 N Montecatina Dr., Tucson, Arizona, 85704, USA\\ \authormark{3}Department of Physics and Material Sciences Center, Philipps-Universit\"at Marburg, Renthof~5, 35032 Marburg, Germany\\ \authormark{4}Department of Mathematics, University of Arizona, 617 N. Santa Rita Ave., Tucson, Arizona 85721, USA}

\email{\authormark{*}ikilen@optics.arizona.edu} 


\begin{abstract}
Microscopic many-body theory coupled to Maxwell’s equation is used to investigate dual-wavelength operation in vertical external-cavity surface-emitting lasers. The intrinsically dynamic nature of coexisting emission wavelengths in semiconductor lasers is associated with characteristic non-equilibrium carrier dynamics which causes significant deformations of the quasi-equilibrium gain and carrier inversion. Extended numerical simulations are employed to efficiently investigate the parameter space to identify the regime for two-wavelength operation. Using an frequency selective intracavity etalon, two families of modes are stabilized with dynamical interchange of the strongest emission peaks. For this operation mode, anti-correlated intensity noise is observed in agreement with the experiment. A method using effective frequency selective filtering is suggested for stabilization genuine dual-wavelength output.
\end{abstract}

\section{Introduction}
Vertical external-cavity surface-emitting lasers (VECSELs) are wavelength tunable solid-state lasers that have been configured to produce either high-power/ultrashort mode-locked pulses or single-/multi-wavelength output \cite{KellerTropper06, tropper2012ultrafast, tilma2015recent, rahimi2016recent, ElectronLett12, LaserPhotonRev12, zhang201423, waldburger2016high, klopp13}. Currently, the peak power produced in multi-mode operation is \unit[106]{W} \cite{ElectronLett12,LaserPhotonRev12} and \unit[23]{W} for single-mode continuous wave output \cite{zhang201423}. VECSELs have been able to generate intracavity mode-locked single pulses of temporal duration around \unit[100]{fs} \cite{waldburger2016high, klopp13, laurain2018modeling}, with the shortest intracavity pulse length at \unit[95]{fs} \cite{laurain2018modeling}. A further unique aspect of these surface emitting lasers is that the single pass gain is low due to light amplification through a periodic or quasi-periodic stacking arrangement  of thin (typically \unit[8-10]{nm} thick) quantum wells (QWs). Consequently, one needs high mirror reflectivities to initiate lasing, meaning that they need to be strongly optically pumped and can sustain very strong intra-cavity circulating fields. As a result these are ideal semiconductor laser systems in which to explore strongly non-equilibrium many-body effects where carrier distributions (electrons, holes, phonons) are strongly perturbed from the standard quasi-equilibrium Fermi distributions.

For a wide range of applications, it is desirable to operate VECSELs in a regime that allows for stable two-wavelength emission. Relevant examples include spectroscopy \cite{axner1987detection, zhou1993terahertz}, differential absorption LIDAR with multiple applications from detection of geological fault lines to measuring the atmospheric $\mathrm{CO}_2$ concentration \cite{langbein1990variations, koch2004coherent}, semiconductor laser parameter measurement \cite{jiang1993parameter, liu1994four}, THz-imaging in contexts ranging from security applications to detection of brain cancer cells \cite{kleine2001continuous, hu1995imaging, yamaguchi2016brain, federici2005thz}, and THz signal generation \cite{wang1995tunable, brown1995photomixing}. In VECSELs THz output has been produced using both intracavity difference frequency generation (DFG) and mode-locked pulses centered at different frequencies with THz spacing \cite{scheller2016dual, scheller2010room}.

In the past, two-wavelength operation in semiconductor lasers has been studied both experimentally and theoretically \cite{leinonen2007dual, pal2010measurement, scheller2010room, chernikov2012time, wichmann2013evolution, wichmann2015antiphase, scheller2017high, de2013intensity, koryukin2007antiphase, ahmed2002influence, yacomotti2004dynamics, matus2004dynamics, baumner2011non}. Stable dual-wavelength operation of VECSELs has been experimentally realized by multiple groups \cite{leinonen2007dual, pal2010measurement, scheller2010room, wichmann2013evolution, wichmann2015antiphase, scheller2017high, chernikov2012time}. In particular, experiments show that there is persistent anti-correlated intensity dependent noise in the dual-wavelength output spectra \cite{wichmann2015antiphase, scheller2017high}. Moustafa Ahmed and Minoru Yamada investigated multi-mode operation in AlGaAs–GaAs and InGaAsP–InP laser systems using multi-mode rate equations coupled to Langevin noise sources \cite{ahmed2002influence}. They found multiple different regimes with stable and unstable single- and multi-mode operation, which they characterize according to injection current and the linewidth enhancement factor. In Ref.~\cite{yacomotti2004dynamics}, Yacomotti et al. have investigated multi-longitudinal-mode operation in semiconductor lasers experimentally and numerically using rate equations for the optically active gain medium. They observed strong anti-correlated mode dynamics between two modes, while the laser emitted constant total intensity. They concluded that four-wave mixing was the dominant cause of the observed phenomena.

Most of the theoretical analysis is based on rate equation approaches \cite{de2013intensity, koryukin2007antiphase, ahmed2002influence, yacomotti2004dynamics, matus2004dynamics}, where the optical polarization dynamics has been adiabatically eliminated and the material excitation is purely modeled by dynamically varying carrier densities. As a consequence, the induced electron-hole polarizations in the QWs are not resolved with the consequence that the ultrafast non-equilibrium light-matter coupling dynamics in VECSEL systems cannot be treated in detail. Moreover, these rate equation approaches rely on added parameterization (although well motivated physically) and directly treat the laser gain - the latter by integrating over the non-equilibrium carrier distributions is insensitive to details such as deformations due to carrier scattering and interaction with the lasing field. 

To improve on the rate-equation analysis, B{\"a}umner et al. investigated non-equilibrium carrier dynamics in a simple micro-resonator system under dual-wavelength operation \cite{baumner2011non}. In their paper, the authors simulate the Maxwell semiconductor Bloch equations (MSBE) at the Hartree-Fock level with simple rate approximations for carrier scattering and found that including non-equilibrium carrier dynamics was essential for accurate modeling of single- and dual-wavelength operation.

In this paper, we will expand the dual-mode operation analysis to VECSELs and numerically solve the MSBE with carrier scattering computed on the level of second Born-Markov approximation \cite{haug09, hader2003microscopic}. In the past, we have employed this model to systematically analyze the role of microscopic carrier dynamics in mode-locked pulse operation of VECSELs, see Refs.~\cite{kilen2017non, kilen2018vecsel, kilen2014ultrafast, kilen2016fully}. In mode-locked operation, pulses with temporal duration in the pico- to femtosecond range are common, which significantly limits the light-matter interaction time within the semiconductor quantum wells during which the detailed carrier dynamics has to be numerically evaluated. Ironically in contrast, a dual-wavelength intracavity field is continuously extracting QW carriers from the inverted states that have to be replenished by external pumping into absorbing states. Furthermore, the coexistence of several wavelength fields inside the cavity leads to dynamic sum- and difference frequency generation such that one never reaches a truly stationary regime. Hence, the high dimensional carrier scattering integrals have to be solved for the entire duration of the simulation, which poses a significant computational challenge but is unavoidable due to the strong non-equilibrium dynamics that influences the round-trip amplification of the intracavity field.

\section{Microscopic Theory}
\label{sec:MicroscopicTheory}
\begin{figure}[ht]
\centerline{\includegraphics[width=1.0\linewidth]{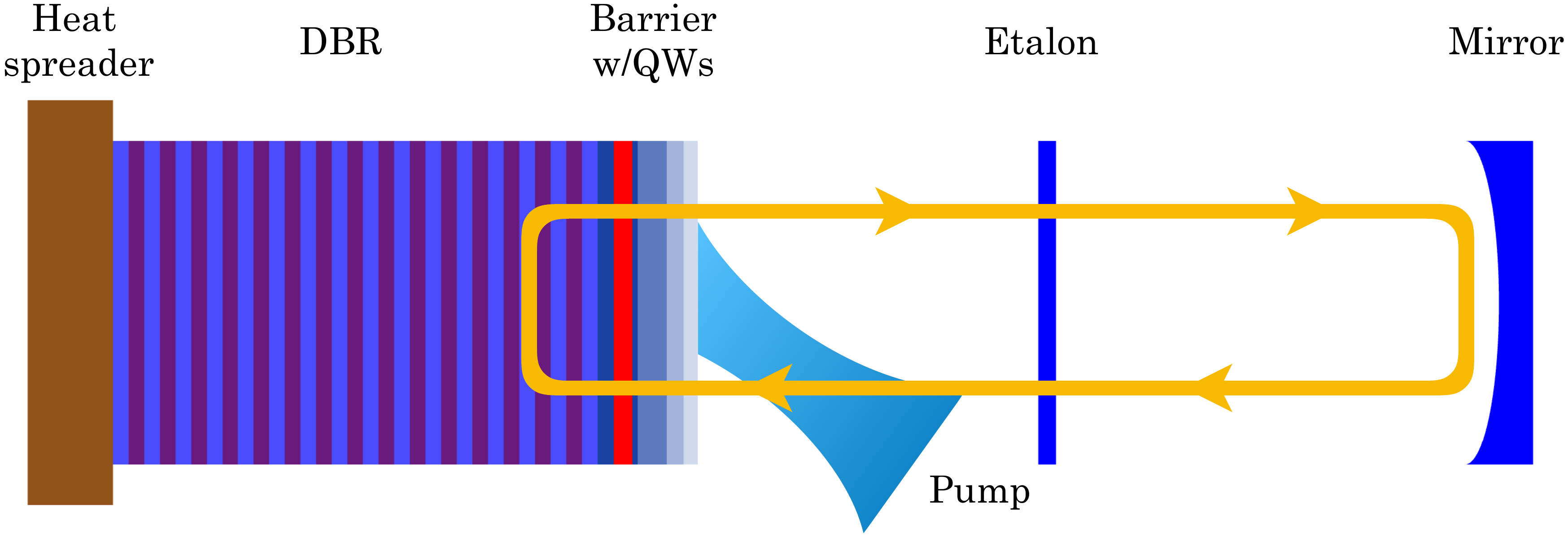}}
\caption{A VECSEL schematic showing a cross section of a linear cavity setup. The light field (yellow arrows) propagates in the external air cavity between the gain chip (left) and an output coupling mirror (right). The optically pumped gain chip consists of: a distributed Bragg reflector (DBR) on top of a heat spreader, a barrier region with optically active QWs (red), a cap layer, and (optionally) a dispersion compensating coating. An etalon can be placed in the cavity to facilitate dual-wavelength operation.}
\label{fig:cavity}
\end{figure}
In Fig.~\ref{fig:cavity}, we present a schematic outline of a VECSEL structure including the QWs, the output coupling mirror and an additional intra-cavity etalon that might be introduced in order to select a specific modal output. The microscopic model of such a VECSEL must fully resolve the time dynamics of the electromagnetic cavity field and the microscopic QW carrier dynamics, where the majority of computational complexity originates from the latter. In order to simplify the numerical challenge, our analysis considers only the propagating light field along the z-axis perpendicular to the QW planes. The electromagnetic field, $E(z,t)$, inside the VECSEL cavity is modeled using Maxwell's equation, 
\begin{equation}
\left[ \frac{\partial^2}{\partial z^2} - \frac{n(z)^2}{c^2_0} \frac{\partial^2}{\partial t^2} \right] E(z,t) = \mu_0 \frac{\partial^2}{\partial t^2} P(z,t)\,.
\label{eq:maxwell}
\end{equation}
Here, $c_0$ is the speed of light in vacuum, $\mu_0$ is the vacuum permeability, $n(z)$ is the material refractive index, $P(z,t)$ is the QW macroscopic polarization. A standard VECSEL cavity cross-section consists of multiple material layers such as: GaAs, AlAs, and an external air cavity, as shown in Fig.~\ref{fig:cavity}. We assume a constant index of refraction for each material layer, which allows us to simulate the propagating field by solving Eq.~(\ref{eq:maxwell}) inside each material layer and couple the layers together using standard Maxwell boundary conditions \cite{baumner2011non}. At the boundary of two different materials, the propagating field will experience reflection and transmission and, most importantly, the field can be amplified or absorbed by interacting with the dynamically changing QW carriers through the macroscopic polarization. 

An optically active QW can be modeled using the multiband semiconductor Bloch equations (SBE) \cite{haug09}, which describe the time evolution of the microscopic electron (hole) occupation numbers, $n^{\mathrm{e(h)}}_{\lambda (\nu), \textbf{k}}$, and the microscopic polarizations, $p_{\lambda,\nu,\textbf{k}}$, for a given crystal momentum \textbf{k} in the conduction (valence) band denoted by $\lambda$ ($\nu$)
\begin{eqnarray}
\label{eq:sbe}
\frac{d}{dt} p_{\lambda,\nu,\textbf{k}} &=& -\frac{\mathrm{i}}{\hbar}\sum_{\lambda_1,\nu_1} \left( e^\mathrm{e}_{\lambda, \lambda_1, \textbf{k}} \delta_{\nu,\nu_1} + e^\mathrm{h}_{\nu, \nu_1, \textbf{k}} \delta_{\lambda,\lambda_1} \right)p_{\lambda_1,\nu_1,\textbf{k}} + \Gamma_{\lambda,\nu,\text{deph}}\nonumber\\
& &- \mathrm{i} \left( n^{\mathrm{e}}_{\lambda, \textbf{k}} + n^{\mathrm{h}}_{\nu, \textbf{k}} -1  \right) \Omega_{\lambda,\nu,\textbf{k}} + \Gamma^{\mathrm{p}}_{\lambda(\nu),\text{spont}}\,,\\
\frac{d}{dt} n^{\mathrm{e(h)}}_{\lambda (\nu), \textbf{k}} &=& -2\, \text{Im}\left(\Omega_{\lambda,\nu,\textbf{k}} (p_{\lambda,\nu,\textbf{k}})^*\right) + \Gamma^{\mathrm{e(h)}}_{\lambda(\nu),\text{scatt}} +\Gamma^{\mathrm{e(h)}}_{\lambda(\nu),\text{spont}} + \left. \frac{d}{dt}n^{\mathrm{e(h)}}_{\lambda(\nu),\textbf{k}}\right|_{cc+cp}\,.\nonumber
\end{eqnarray}
The macroscopic polarization, that feeds back into Eq.~(\ref{eq:maxwell}), is defined by $P(z,t) = \sum_{\lambda,\nu,\textbf{k}} d^\textbf{k}_{\lambda \nu} p_{\lambda,\nu,\textbf{k}}$, where $d^\textbf{k}_{\lambda \nu}$ is the dipole matrix element. The single particle electron (hole) energies are given by $\epsilon^{\mathrm{e(h)},\lambda(\nu)}_{\textbf{k}}$ and couple through the Coulomb potential, $V^{\lambda,\nu_1,\nu,\lambda_1}_{|\textbf{k}-\textbf{q}|}$. The Hartree-Fock renormalized single particle energies are
\begin{equation}
\begin{split}
e^{\mathrm{e}}_{\lambda, \lambda_1, \textbf{k}} & = \epsilon^{\mathrm{e},\lambda}_{\textbf{k}}\delta_{\lambda,\lambda_1} - \sum_{\lambda_2, \textbf{q}} V^{\lambda,\lambda_2,\lambda_1,\lambda_2}_{|\textbf{k}-\textbf{q}|} n^\mathrm{e}_{\lambda_2,\textbf{q}}
\,,\\
e^{\mathrm{h}}_{\nu, \nu_1, \textbf{k}} & = \epsilon^{\mathrm{h},\nu}_{\textbf{k}}\delta_{\nu,\nu_1} - \sum_{\nu_2, \textbf{q}} V^{\nu,\nu_2,\nu_1,\nu_2}_{|\textbf{k}-\textbf{q}|} n^\mathrm{h}_{\nu_2,\textbf{q}}\,,\\
\end{split}
\end{equation}
and the propagating electromagnetic field is a source in Eq.~(\ref{eq:sbe}) through the effective Rabi frequency, $\Omega_{\lambda,\nu,\textbf{k}}$, defined by
\begin{equation}
\begin{split}
\hbar \Omega_{\lambda,\nu,\textbf{k}} & = d^{\lambda,\nu}_{\textbf{k}} E(z,t) + \sum_{\lambda_1,\nu_1,\textbf{q}\neq\textbf{k}} V^{\lambda,\nu_1,\nu,\lambda_1}_{|\textbf{k}-\textbf{q}|} p_{\lambda_1,\nu_1,\textbf{q}}\,.
\end{split}
\end{equation}
In Eq.~(\ref{eq:sbe}), the higher order correlation contributions give rise to the polarization dephasing ($\Gamma_{\lambda,\nu,\text{deph}}$), the carrier-carrier and carrier-phonon scattering ($\left. \frac{d}{dt}n^{\mathrm{e(h)}}_{\lambda(\nu),\textbf{k}}\right|_{cc+cp}$), the carrier screening, and the carrier replenishing ($\Gamma^{\mathrm{e(h)}}_{\lambda(\nu),\text{scatt}}$). In addition, the semiclassical model includes spontaneous emission though, $\Gamma^{\mathrm{e(h)}}_{\lambda(\nu),\text{spont}}$ and $\Gamma^{\mathrm{p}}_{\lambda(\nu),\text{spont}}$. The carrier scattering is of particular interest and is split into the carrier-carrier ($\left.\frac{d}{dt}n^{\mathrm{e}}_{\lambda, \textbf{k}}\right|_\mathrm{cc}$) and carrier-phonon ($\left.\frac{d}{dt}n^{\mathrm{e}}_{\lambda, \textbf{k}}\right|_\mathrm{cp}$) scattering shown for electrons in Eqs.~(\ref{eq:sbe_cc}-\ref{eq:sbe_cp}), as computed by Hader et al. in Ref.~\cite{hader2003microscopic},
\begin{eqnarray}
\label{eq:sbe_cc}
\left.\frac{d}{dt}n^{\mathrm{e}}_{\lambda, \textbf{k}}\right|_\mathrm{cc} & = & \sum_{\substack{ \lambda_1,\lambda_2,\lambda_3\\ \textbf{q},\textbf{k}' }} \hat{V}^{\lambda,\lambda_1,\lambda_2,\lambda_3}_{|\textbf{q}|}\left(\hat{V}^{\lambda,\lambda_1,\lambda_2,\lambda_3}_{|\textbf{q}|} - \hat{V}^{\lambda,\lambda_1,\lambda_2,\lambda_3}_{|\textbf{k}-\textbf{k}'|}\right) \mathcal{D} \left(\epsilon^{\mathrm{e},\lambda}_{\textbf{k}} + \epsilon^{\mathrm{e},\lambda_1}_{\textbf{k}'-\textbf{q}} - \epsilon^{\mathrm{e},\lambda_2}_{\textbf{k}'} - \epsilon^{\mathrm{e},\lambda_3}_{\textbf{k}-\textbf{q}}\right)\cdot \nonumber \\
& & \left( (1-n^{\mathrm{e}}_{\lambda_2, \textbf{k}'})(1-n^{\mathrm{e}}_{\lambda_3, \textbf{k}-\textbf{q}}) n^{\mathrm{e}}_{\lambda, \textbf{k}} n^{\mathrm{e}}_{\lambda_1, \textbf{k}'-\textbf{q}} - (1-n^{\mathrm{e}}_{\lambda, \textbf{k}})(1-n^{\mathrm{e}}_{\lambda_1, \textbf{k}'-\textbf{q}})n^{\mathrm{e}}_{\lambda_2, \textbf{k}'} n^{\mathrm{e}}_{\lambda_3, \textbf{k}-\textbf{q}} \right) \nonumber \\
&+& \sum_{\substack{\lambda_1, \nu_1, \nu_2 \\\textbf{q},\textbf{k}'}} \left| \hat{V}^{\lambda,\nu_1,\nu_2,\lambda_1}_{|\textbf{q}|} \right|^2 \mathcal{D} \left(\epsilon^{\mathrm{e},\lambda_1}_{\textbf{k}-\textbf{q}} - \epsilon^{\mathrm{e},\lambda}_{\textbf{k}} + \epsilon^{\mathrm{h},\nu_1}_{\textbf{k}'+\textbf{q}} - \epsilon^{\mathrm{h},\nu_2}_{\textbf{k}'}  \right) \cdot  \\
& & \left( (1-n^{\mathrm{e}}_{\lambda_1, \textbf{k}-\textbf{q}})(1-n^{\mathrm{h}}_{\nu_1, \textbf{k}'+\textbf{q}})n^{\mathrm{e}}_{\lambda, \textbf{k}} n^{\mathrm{h}}_{\nu_2, \textbf{k}'} - (1-n^{\mathrm{e}}_{\lambda, \textbf{k}})(1-n^{\mathrm{h}}_{\nu_2, \textbf{k}'}) n^{\mathrm{e}}_{\lambda_1, \textbf{k}-\textbf{q}} n^{\mathrm{h}}_{\nu_1, \textbf{k}'+\textbf{q}} \right).\nonumber
\end{eqnarray}
Here, $\hat{V}^{\lambda,\nu_1,\nu_2,\lambda_1}_{|\textbf{q}|}$ is the screened multiband Coulomb matrix element and $\mathcal{D}(E)$ is an energy conserving function that evaluates to unity for allowed transitions and zero otherwise. The carrier-phonon scattering is evaluated from
\begin{eqnarray}
\label{eq:sbe_cp}
\left.\frac{d}{dt}n^{\mathrm{e}}_{\lambda, \textbf{k}}\right|_\mathrm{cp} & = & \sum_{\lambda_1,\textbf{q}} \left| g^{\lambda \lambda_1}_\textbf{q} \right|^2 \left\{ \mathcal{D}\left( \epsilon^{\mathrm{e},\lambda}_\textbf{k} - \epsilon^{\mathrm{e},\lambda_1}_{\textbf{k}+\textbf{q}} + \hbar \omega_\textbf{q} \right) \left( s_\textbf{q} n^{\mathrm{e},\lambda}_\textbf{k} \left(1-n^{\mathrm{e},\lambda_1}_{\textbf{k}+\textbf{q}} \right) - (1+s_\textbf{q})n^{\mathrm{e},\lambda_1}_{\textbf{k}+\textbf{q}} (1-n^{\mathrm{e},\lambda}_\textbf{k}) \right) \right. \nonumber \\
& -& \left. \mathcal{D}\left( \epsilon^{\mathrm{e}}_{\lambda_1,\textbf{k}-\textbf{q}} - \epsilon^{\mathrm{e}}_{\lambda,\textbf{k}} + \hbar \omega_\textbf{q} \right)\left( s_\textbf{q} n^{\mathrm{e}}_{\lambda_1,\textbf{k}-\textbf{q}} \left(1-n^{\mathrm{e}}_{\lambda,\textbf{k}} \right) - (1+s_\textbf{q})n^{\mathrm{e}}_{\lambda,\textbf{k}} (1-n^{\mathrm{e}}_{\lambda_1,\textbf{k}-\textbf{q}}) \right)\right\}.
\end{eqnarray}
Where $g^{\lambda \lambda_1}_\textbf{q}$ is the Fr{\"o}hlich matrix element \cite{kuhn1998density, waldmuller2005intersubband}, $s_\textbf{q}$ is the phonon distribution, $\hbar \omega_\textbf{q}$ is the LO-phonon energy. The phonon scattering is defined by specifying the lattice parameters: the LO-phonon energy ($\hbar \omega_{LO}=\,$\unit[33.95]{meV}), lattice temperature (\unit[300]{K}) and the dielectric constants at high- ($\epsilon_\infty=\,$11.24) and low- ($\epsilon=\,$13.47) frequencies. These values are determined from interpolation of table values found in Ref.~\cite{landolt1987numerical}.

In order to simplify the numerical challenge, we assume that the relevant portion of the detailed semiconductor bandstructure can be approximated by a two-band effective mass model with transition energy given by $\hbar \omega_{\textbf{k}} = E_\mathrm{g} + \frac{\hbar^2 \textbf{k}^2}{2 m_\mathrm{e}} + \frac{\hbar^2 \textbf{k}^2}{2 m_\mathrm{h}}$. Here, $E_\mathrm{g}$ is the band gap and $m_{\mathrm{e(h)}}$ denotes the effective electron (hole) mass. Furthermore, with the high QW carrier density needed for population inversion, the polarization dephasing is approximated using a characteristic timescale, $\tau_\text{deph}$, such that $\Gamma_{\text{cv},\text{deph}} = -(1/\tau_\text{deph})p_{\text{cv},\textbf{k}}$. In the dephasing time approximation, carrier screening will be treated using the static Lindhard formula \cite{haug09}. With two parabolic bands, Eqs.~(\ref{eq:sbe_cc}-\ref{eq:sbe_cp}) can be further simplified by using the analytic band structure to explicitly evaluate the innermost integral over the energy conserving delta function. Employing a delta function is particularly useful for carrier-carrier scattering because it reduces the number of integrals by one and allows for a considerable computational speedup. 

Our simulation domain consists of a gain chip with ten QWs, an air gap, and an output coupling mirror. This gain chip consists of a GaAs/AlGaAs distributed Bragg reflector, a barrier region with \unit[8]{nm} InGaAs QWs arranged on an anti-node of the DBR standing wave, a cap layer, and an anti-reflection coating. The gain chip is optimized for ultrashort pulse generation with a peak gain at \unit[1030]{nm}, similar to the one used for Ref.~\cite{kilen2017non}. However, mode-locked considerations turn out to be of less importance in the context of dual-wavelength generation. 

The QW carrier replenishing is modeled with a relaxation to a background Fermi distribution, $f^{\mathrm{e(h)}}_{\text{c}(\text{v}), \textbf{k}}$, at a fixed density (\unit[$3.0\cdot 10^{16}$]{$\text{m}^{-2}$}) and temperature (\unit[300]{K}) with $\Gamma^{\mathrm{e(h)}}_{\text{c}(\text{v}),\text{scatt}} = -(n^{\mathrm{e(h)}}_{\text{c}(\text{v}), \textbf{k}} - f^{\mathrm{e(h)}}_{\text{c}(\text{v}), \textbf{k}}) /\tau_{\text{scatt}}$. The characteristic timescale for gain recovery is on the order of tens of picoseconds or longer, which is orders of magnitudes slower than the carrier scattering \cite{hader2016ultrafast}. Numerical tests show that the details of the results are not critically sensitive on the exact choice of $\tau_{\text{scatt}}$. For the studies in this paper, we use $\tau_{\text{scatt}}= 30\,\text{ps}$ which, for appropriately chosen output coupling losses, results in reasonably high intracavity field intensities that are, however, still well below the CW damage threshold for GaAs \cite{smith1972surface, sam1973laser, qi2008investigation}. 

The spontaneous emission of photons into the lasing cavity is approximated using $\Gamma^{\mathrm{e(h)}}_{\lambda(\nu),\text{spont}}= -\Gamma^{\text{spont}}_{\textbf{k}} n^e_{\textbf{k}} n^h_{\textbf{k}}$ and $\Gamma^{\mathrm{p}}_{\lambda(\nu),\text{spont}}= \beta \Gamma^{\text{spont}}_{\textbf{k}} n^e_{\textbf{k}} n^h_{\textbf{k}}$, as used by  B{\"a}umner et al. in Ref.~\cite{baumner2011non}. Here,  $\Gamma^{\text{spont}}_{\textbf{k}} = \frac{n^3_{\mathrm{bgr}}}{\pi^2 \epsilon \hbar^4 c^3_0} |d^{\text{cv}}_{\textbf{k}}|^2 \left( E_g + \frac{\hbar^2 \textbf{k}^2}{2 \text{m}_r} \right)^3$, $\beta$ is a complex number with a random phase factor, and $\epsilon$ is the permittivity. The coupling of light from the QW into the cavity is given by $|\beta|$, and we will for simplicity assume that $|\beta|=1$. The presence of noise in the cavity does not significantly change results and will automatically eliminate any dual-wavelength solutions that are only marginally stable.

\section{Numerical results and discussion}

In order to identify situations where a semiconductor laser operates under stable two-wavelength emission, we have to find conditions where (i) the desired two modes experience comparable round-trip amplification and (ii) where all the other possible modes, in particular those that see more gain, are effectively suppressed. 

\subsection{Dual-wavelength amplification}
\label{sec:dwa}

To realize the equal amplification configuration, one is tempted to look at the quasi-equilibrium carrier gain of the active QWs. However, as it turns out, this is not very useful because of the strong non-equilibrium effects introduced by the self-consistent coupling of the propagating intracavity field with the electronic QW excitations.   

\begin{figure}[ht]
\centerline{\includegraphics[width=1.0\linewidth]{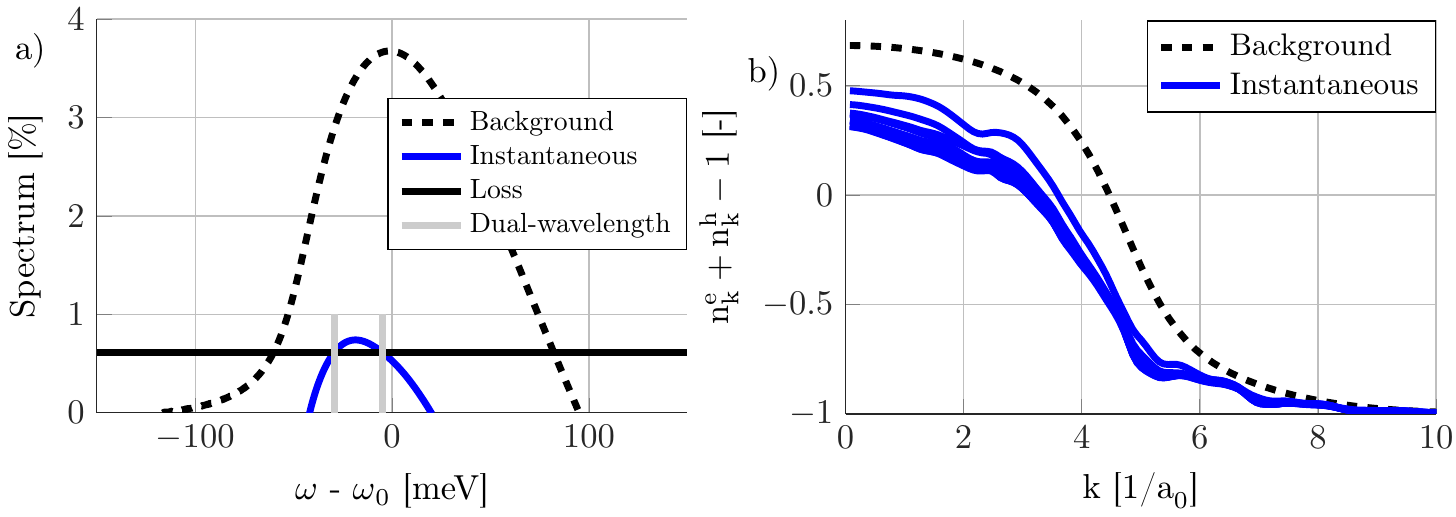}}
\caption{a) The linear QW gain computed without an intracavity field (dashed black) and the non-equilibrium round-trip amplification during dual-wavelength operation (blue). The solid black lines indicates the output coupling loss and the vertical grey lines the spectral location of the dual-wavelength output. b) The family of blue curves depict a time snapshot of the non-equilibrium carrier inversion of each QW in the gain medium. The black dashed curve depicts carriers in a background quasi-equilibrium distributions.}
\label{fig:nonEqGain}
\end{figure}
When a laser is switched on, we have to provide a sufficiently high carrier density to allow for QW inversion, to have the gain needed for round-trip amplification. The intracavity field that builds up due to spontaneous emissions from the inverted QWs will be constructively amplified at the resonant cavity modes and the laser output stabilizes once the intracavity field intensity saturates the gain. The intensity of each spectral peak depends on the transient dynamics during buildup from noise. Under quasi-stationary operation conditions, the carrier density and the carrier distributions are dynamically established by the balancing process of competing momentum selective carrier removal through the intracavity field and the carrier replenishment by the pump process. As a consequence, the carrier density and distribution and thus the instantaneous gain significantly deviate from the quasi-equilibrium conditions. An example for such a situation in shown in Fig.~\ref{fig:nonEqGain}b). Fig.~\ref{fig:nonEqGain}b) shows low-momentum \emph{kinetic holes} burned in the carrier distributions, where dual-wavelength lasing photons are being extracted, and high-momentum phonon scattering induced distortions near LO-phonon resonances (see more in section \ref{sec:sc_disc}). 

Using the standard gain approach, for dual-wavelength operation to occur, the instantaneous gain for each running mode has to compensate the respective losses, thus clamping the {\it non-equilibrium} gain at these frequencies. An example for such a stable dual-wavelength situation is  shown Fig.~\ref{fig:nonEqGain}a) where the non-equilibrium gain (blue) has clamped to the loss line (solid black). The vertical gray lines depict the dual-wavelengths. As discussed above, Fig.~\ref{fig:nonEqGain}b) depicts the corresponding non-equilibrium inversion in the individual QWs that is responsible for the round-trip amplification. Comparing the actual and the quasi-equilibrium configurations, we see that that the quasi-equilibrium gain and inversion have limited use in predicting the conditions for dual-wavelength operation.

In order to evolve the full field within the VECSEL cavity we have to solve the full coupled MSBE~(\ref{eq:maxwell}-\ref{eq:sbe}) numerically where we account for the optical pumping that continuously generates carriers in high-momentum states whereas the lasing emission selectively removes carriers from specific regions of low-momentum states. This carrier replenishing dynamics is governed by the carrier Coulomb and phonon scattering such that the time-resolved ultrafast QW carrier dynamics forces the timesteps on the order of a fraction of a femtosecond. On a longer timescale, the convergence of the intracavity field to a stable solution takes several thousand cavity round-trips. In contrast to mode-locked pulse solutions with 100's of femtosecond duration (e.g., see Refs.~\cite{kilen2018vecsel} and \cite{kilen2016fully}), the intracavity dual-wavelength field is always interacting with the optically active QWs. This continually drives the microscopic QW carriers into dynamic non-equilibrium distributions, which are never truly stationary due to intrinsic multi-mode field dynamics e.g. in the beating with the sum- and difference-frequency field components.

The need to include the full microscopic carrier scattering via Eqs.~(\ref{eq:sbe_cc}-\ref{eq:sbe_cp}) to arrive at an asymptotic non-equilibrium solution of the coupled MSBE poses a significant numerical challenge. The multiple dimensional integrals contain millions of multiplications, which effectively increases the computational demand of each time step by orders of magnitude relative to scattering-rate approximations. Luckily, by taking advantage of modern parallel supercomputers, and the implementation of efficient parallelization strategies we can make headway on these problems. We run simulations on an SGI UV2000 computer with 384 cores and 4TB of shared memory, which enables us to run multiple simulations in parallel. Each simulation can take from \unit[12]{hours} up to \unit[30]{days} to determine the intracavity field stability.

Given the complexity of these full simulations it is imperative that we devise a robust and efficient method to a priori find promising conditions for initializing the simulation with the goal of finding potentially stable dual-wavelength emission. In order to efficiently determine the regime of potentially stable two-wavelength operation, we have devised a mapping method to initialize the system close to a stable solution with two equal spectral intensities. For this purpose, we specify a gain chip with QWs initialized at a chosen background density. For each pair of frequencies $(\omega_1, \omega_2)$, we inject an ideal dual-wavelength field of the form, $E(t) = E_\text{in} (\mathrm{e}^{-\mathrm{i} \omega_1 t} + \mathrm{e}^{-\mathrm{i} \omega_2 t})$, until the asymptotic carrier distributions stabilize, i.e., the gain chip gives constant spectral amplification. At this point, the \emph{injection} amplification of both frequencies are recorded, which are not necessarily equal. We then repeat this process for multiple input field amplitudes, $E_\text{in}$, until an amplitude that results in \emph{equal} spectral amplification of both frequencies is found. However, we note that there are frequency pairs where no field amplitude will give equal spectral amplification of modes.

\begin{figure}[ht]
\centerline{\includegraphics[width=1.0\linewidth]{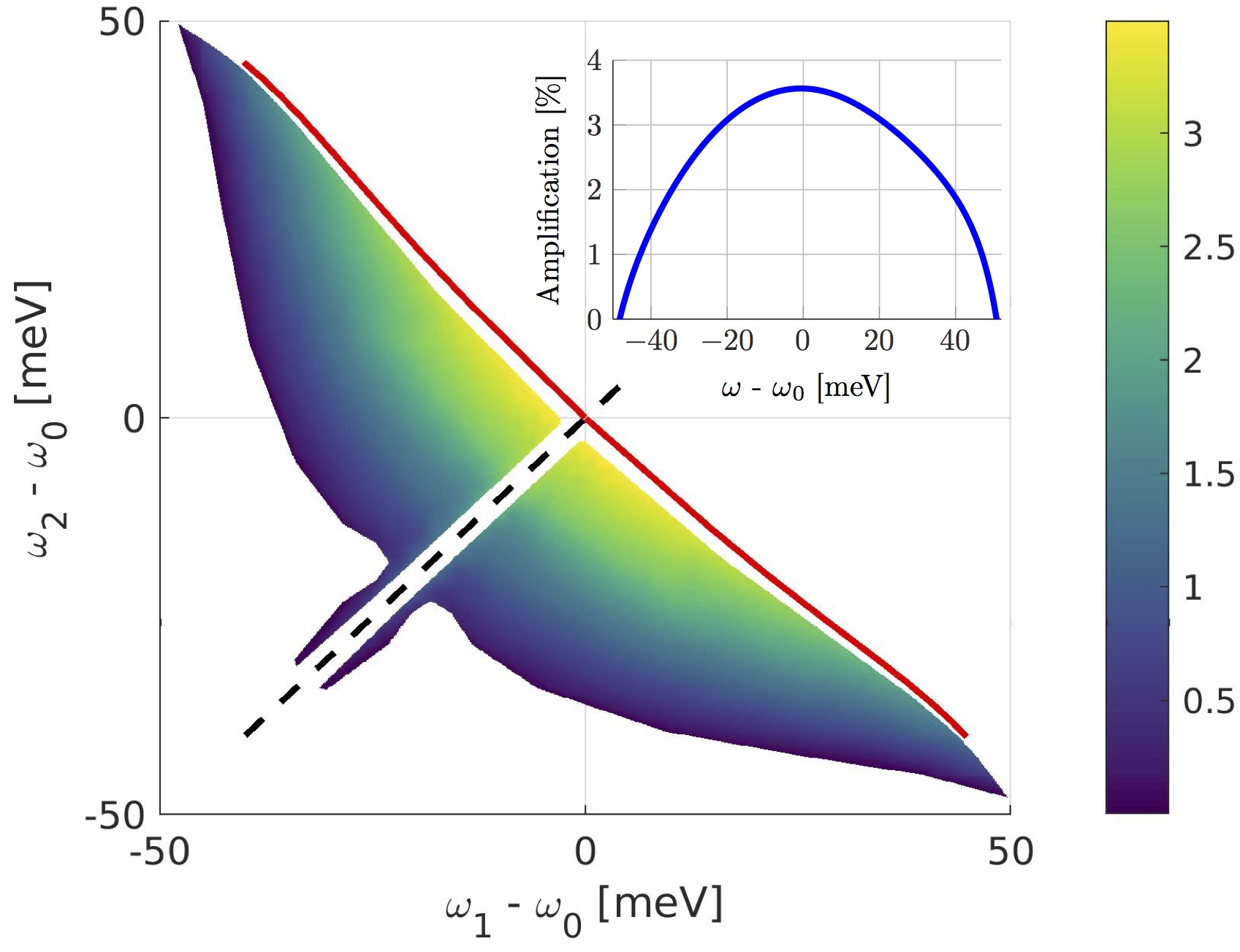}}
\caption{Overview of asymptotic dual-wavelength solutions (color surface) for frequencies ($\omega_1$, $\omega_2$) relative to peak gain computed with the injection mapping scheme for a specific gain chip. The surface amplitude gives the non-equilibrium round-trip amplification that a dual-wavelength solution with the given frequencies would experience from the gain chip. The solid red line indicates the two intersections of a constant loss and the  background round-trip amplification, shown as an inset. Single frequency operation ($\omega_1=\omega_2$) is indicated with a dashed black line.}
\label{fig:surface}
\end{figure}
The resulting dual-wavelength field with equal amplification for each pair of frequencies is then an idealized approximation of the real situation because there is no feedback, spectral broadening, phase difference between the peaks, nonlinear wave mixing, or frequency drift included in the solution. To obtain a self-consistent solution of the full problem, we use these approximate solutions as input fields. In this way, we can very efficiently limit the parameter space where we find stable dual-wavelength output and the reduced computational time allows us to study the full non-equilibrium carrier dynamics. 

Fig.~\ref{fig:surface}, summarizes the outcome of applying the mapping technique for various choices of injected dual-wavelengths. The color surface shows the amplification that potentially stable dual-wavelength solutions experience during a single pass through the gain chip. Points outside of this surface does not have a dual-wavelength solution with equal amplification of the spectral peaks. However, this would not rule out multiple families of stable unequal amplitude dual-wavelength solutions.

The red line in Fig.~\ref{fig:surface} is obtained by recording the intersections of the background amplification with a constant loss. The background amplification, shown as an inset in Fig.~\ref{fig:surface}, is the undisturbed QW gain modulated by the DBR and other material layers. This red line represents the theoretical limit of dual-wavelength operation where the QW carriers would be equal to the background, i.e., undisturbed pump density. Indeed, the background round-trip amplification approximates the amplification of dual-wavelength solutions along the nearby surface edge.

\subsection{Dual-wavelength stabilization}
Classically, gain clamping to the cavity loss will dictate single-wavelength output from a laser. Hence, in order to establish dual-wavelength operation, it is necessary to suppress emission at the peak gain wavelength by restricting the possible lasing modes. This suppression can be accomplished using either frequency selective gratings, etalons, or a sufficiently short optical cavity with properly placed longitudinal modes.

The first method involves a carefully selected external cavity length, which will ensure that the longitudinal cavity modes overlap with the two chosen wavelengths and no other mode experiences higher gain. The possible cavity lengths are usually short (less than a hundred micrometers) and result in strict spectral filtering because only the resonant cavity modes will be amplified. The second method involves inserting an additional etalon inside the cavity. In the third method, a finite impulse response (FIR) frequency filter is inserted into the cavity.

\subsubsection{Short cavity}
\label{sec:sc_disc}
\begin{figure}[ht]
\centerline{\includegraphics[width=1.0\linewidth]{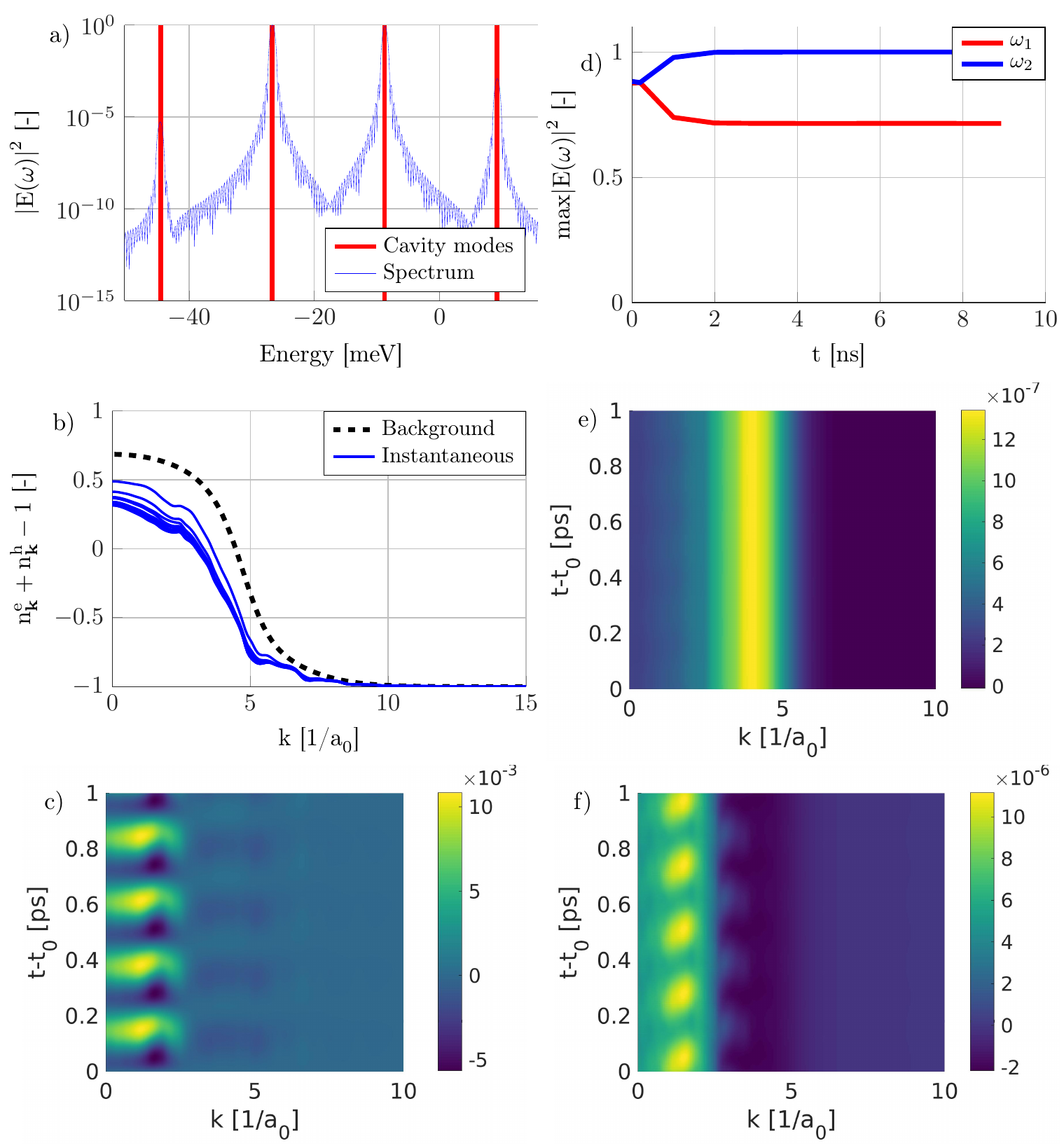}}
\caption{An example simulation with stable dual-wavelength operation using the fully microscopic theory. a) The resulting stable output spectra is shown, in blue, with the cavity frequency filtering in red. b) The inversion in the QWs during stable operation. c) Change in carrier inversion relative to $t=t_0=$\unit[9]{ns}. d) The relative intensity of the two largest spectral peaks. e) Number of electrons replenished from pumping, $\Gamma^{\mathrm{e}}_{\text{scatt}}$. f) Change in electrons from the sum of carrier-carrier and carrier-phonon scattering, $\left. \frac{d}{dt}n^{\mathrm{e}}_{\textbf{k}}\right|_{cc+cp}$.}
\label{fig:solutions}
\end{figure}

To illustrate the dual-mode dynamics, we start with a very short-cavity configuration since this provides the most straightforward mode selection method. Fig.~\ref{fig:solutions} gives an overview of the cavity field and microscopic dynamics for a cavity of length \unit[31]{$\mu$m} resulting in frequencies at ($\omega_1$, $\omega_2$) = (\unit[-26.7]{meV}, \unit[-8.8]{meV}), i.e., \unit[4.3]{THz} spacing, in Fig.~\ref{fig:surface}. Fig.~\ref{fig:solutions}(a) presents the stable output spectrum whereas (b) shows a snapshot of the inversion in all ten QWs (blue) during stable operation. We notice the pronounced k-dependent differences between the actual and the background inversion (black dashed line). The strongest deformations result from the k-selective carrier removal by the running modes. In addition, we also notice modifications at higher k-values away from the main spectral peaks that originate from the LO-phonon resonances in the carrier-phonon scattering sums, Eq.~\ref{eq:sbe_cp}.

In Fig.~\ref{fig:solutions}(c), we depict the dynamic inversion change for the example of the 6\textsuperscript{th} QW away from the DBR. The dual mode intracavity field is continually extracting carriers. The dynamic mode beating results in a dynamic k-dependent inversion oscillation around an equilibrium value. Even though these oscillation amplitudes are reduced once the intracavity field has stabilized, the carrier distributions never become truly stationary for dual-wavelength output. Fig.~\ref{fig:solutions}(e-f) shows the time dynamics of the electron replenishing and carrier scattering. The carrier pumping relaxes the carriers to the background distribution, shown as a black dashed line in Fig.~\ref{fig:solutions}(b). In this particular example, hot carriers are injected into the QWs with a peak right below $k=5$. Carrier scattering is equilibrating the hot high momentum carriers into the low momentum region, which the cavity field can then extract, see Fig.~\ref{fig:solutions}(f).

\subsubsection{Intracavity etalon}
\begin{figure}[ht]
\centerline{\includegraphics[width=1.0\linewidth]{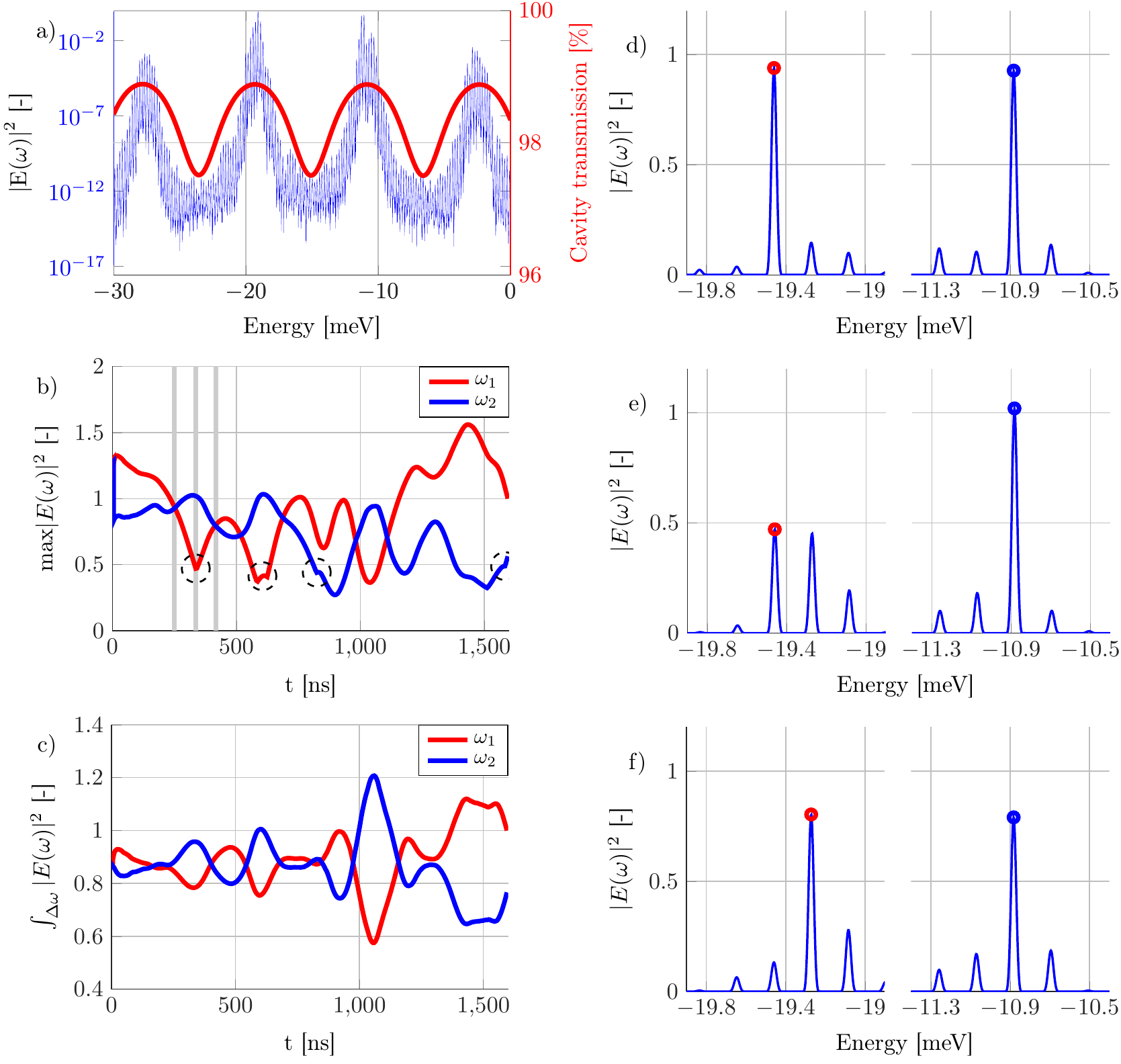}}
\caption{Example simulation using a longer external air cavity (\unit[3.3]{mm}) with a glass etalon. a) The resulting stable output spectra (blue) with the cavity transmission (red). b) The relative intensity of the two largest spectral peaks (red/blue), dashed circles indicates where a nearby mode overtakes the current dominant mode. c) The dynamics of the integrated spectral intensity near \unit[-19.4]{meV} and \unit[-10.9]{meV} in a). d-f) Snapshots of the resulting spectrum at times \unit[250]{ns}, \unit[336]{ns}, and \unit[418]{ns}. The red/blue circles indicate the two dominant modes, as seen in b), and the locations of the snapshots are indicated by thick vertical grey lines in b)}
\label{fig:extraSolutions}
\end{figure}
A common method to generate dual-wavelength operation in experiments is to use an intracavity etalon. To study this scenario, we present in Fig.~\ref{fig:extraSolutions} the results of a simulation with a glass ($n=1.5$) etalon of length \unit[99.1]{$\mu$m} which has been placed inside an external cavity of length \unit[3.3]{mm}. The etalon is configured to induce dual-wavelength output with a frequency spacing of \unit[2]{THz} at (\unit[-19.4]{meV},\unit[-10.9]{meV}), compare Fig.~\ref{fig:surface}. The etalon is placed near the output coupling mirror which results in a modulated output coupling loss. The computed emission spectrum is shown in panel a) superimposed on the cavity transmission. We notice that the output spectrum consists of multiple families of modes located around each transmission peak, which is clearly far from our initial guess of only two spectral peaks but the generic picture in Fig.~\ref{fig:surface} still offers key guidance.

In Fig.~\ref{fig:extraSolutions}b), we see the spectral intensity of the two highest output modes competing for amplification. In particular, the black dashed circles show where an adjacent mode overtakes the momentarily dominant wavelength. The total energy in the mode families surrounding the spectral peaks at \unit[-19.4]{meV} and \unit[-10.9]{meV} are compared in Fig.~\ref{fig:extraSolutions}c), by integrating over a \unit[$\pm$4]{meV} region. The time dynamics clearly shows that the spectral peaks are strongly anti-correlated. Similar behavior has been observed in experiments \cite{wichmann2015antiphase, scheller2017high}.

In panels d-f), we present snapshots of the output spectra at \unit[250]{ns}, \unit[336]{ns}, and \unit[418]{ns} after initialization. The three snapshots show the spectral dynamics surrounding the two main families of modes located around \unit[-19.4]{meV} and \unit[-10.9]{meV}. Starting from Fig.~\ref{fig:extraSolutions}d) at \unit[250]{ns}, we see that the dominant mode near \unit[-19.4]{meV} is overtaken by an adjacent mode around \unit[336]{ns}, shown in Fig.~\ref{fig:extraSolutions}e-f). These snapshots are marked in Fig.~\ref{fig:extraSolutions}b) with thick vertical grey bars and show a typical situation. Indeed, there are similar events indicated by the dashed black circles in Fig.~\ref{fig:extraSolutions}b). Altogether, even though we run the simulations for up to \unit[1600]{ns}, i.e., for a total simulation time corresponding to about \unit[30]{days}, the emission spectra never stabilize due to the ongoing competition between the different modes. This clearly emphasizes the relatively poor performance of the intracavity etalon which not only provides frequency selective filtering but also a certain amount of intra-cavity reflected light that leads to destabilization of genuine two-mode operation. However, in a variety of practical settings (e.g., intracavity DFG THz generation) a clean dual-wavelength THz signal might not be required \cite{scheller2017high}.

\subsubsection{Finite Impulse Response Filter}
In experimental setups, the external cavity can include a frequency selective grating which will spatially separate the intracavity field spectrum \cite{breede2002fourier, wang1994tunable, lee1997stable}. The separated frequencies are reflected by a frequency selective mirror and returned to the cavity. This method allows for a high degree of spectral control in the filtering of the intracavity field. Methodically, this procedure first applies a Fourier transform to the intracavity field and then a spectral filter to the back reflected signal which is returned to the cavity \cite{breede2002fourier}. 

\begin{figure}[ht]
\centerline{\includegraphics[width=1.0\linewidth]{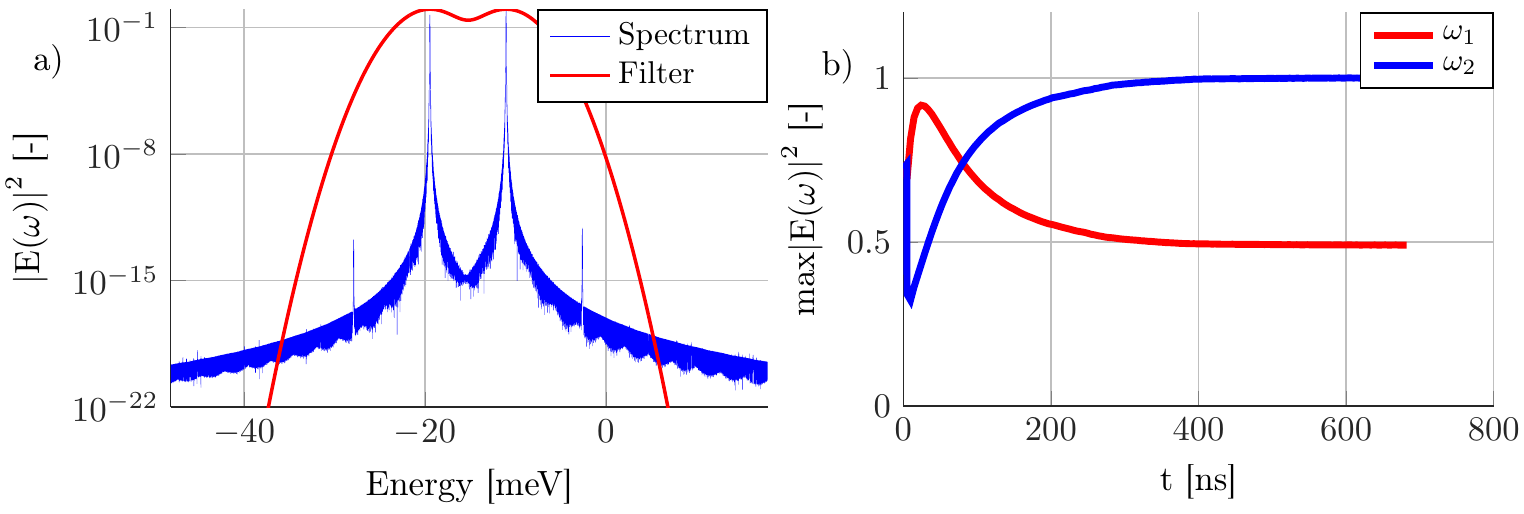}}
\caption{Example simulations of dual-wavelength operation using a longer external air cavity (\unit[3.3]{mm}) with an intracavity frequency filter. a) The spectral shape of the filter is shown as red line and the resulting stable output spectrum is presented as the blue curve. b) The relative intensity of the two largest spectral peaks is shown as function of time after initialization.}
\label{fig:extraSolutions2}
\end{figure}
In our simulations, an idealized version of this method can be implemented through a FIR filter with a customizable filter function. The FIR filter is placed in the external air cavity and provides a frequency dependent loss to the intracavity field through a convolution with the impulse response of a given filter function.

An example the simulation results is shown in Fig.~\ref{fig:extraSolutions2} where we include a FIR frequency filter in a  \unit[3.3]{mm} external cavity with spectral peaks centered at \unit[-19.4]{meV} and \unit[-10.9]{meV}. Here, we have constructed the filter function as two Gaussian peaks with \unit[1]{THz} full width half maximum and \unit[2]{THz} peak-to-peak spacing. The resulting output spectra is shown in Fig.~\ref{fig:extraSolutions2}a) and it is clearly very different from the one produced by the etalon, in Fig.~\ref{fig:extraSolutions}a). As we can see in Fig.~\ref{fig:extraSolutions2}, the FIR filter successfully stabilizes \unit[2]{THz} dual-wavelength operation at (\unit[-19.4]{meV}, \unit[-10.9]{meV}). The FIR filter is also useful to realize other solutions on the surface of Fig.~\ref{fig:surface} with mode separation less than \unit[2]{THz}, which are unreachable with a short external cavity.

\subsection{Self-consistent dual-wavelength results}
To obtain a general overview, we now scan the parameter space for optimal dual mode operation conditions. For this purpose, we perform series of numerical simulations solving the full microscopic equations, always initializing the runs utilizing the results summarized in Fig.~\ref{fig:surface}. As we described earlier in Sec.~\ref{sec:dwa}, we choose a pair of frequencies, $(\omega_1, \omega_2)$, as well as the initial field amplitude, $E_\text{in}$, which gives equal spectral amplification for both frequencies. At first, we introduce the QWs into a stable state and allow the approximate -- not yet self-consistent -- dual-wavelength field to fill the cavity, before we set the output coupling loss equal to the spectral amplification. The full simulation is then started from this initial configuration and is run until the feedback from the cavity equilibrates the system into a self-consistent solution.

\begin{figure}[ht]
\centerline{\includegraphics[width=1.0\linewidth]{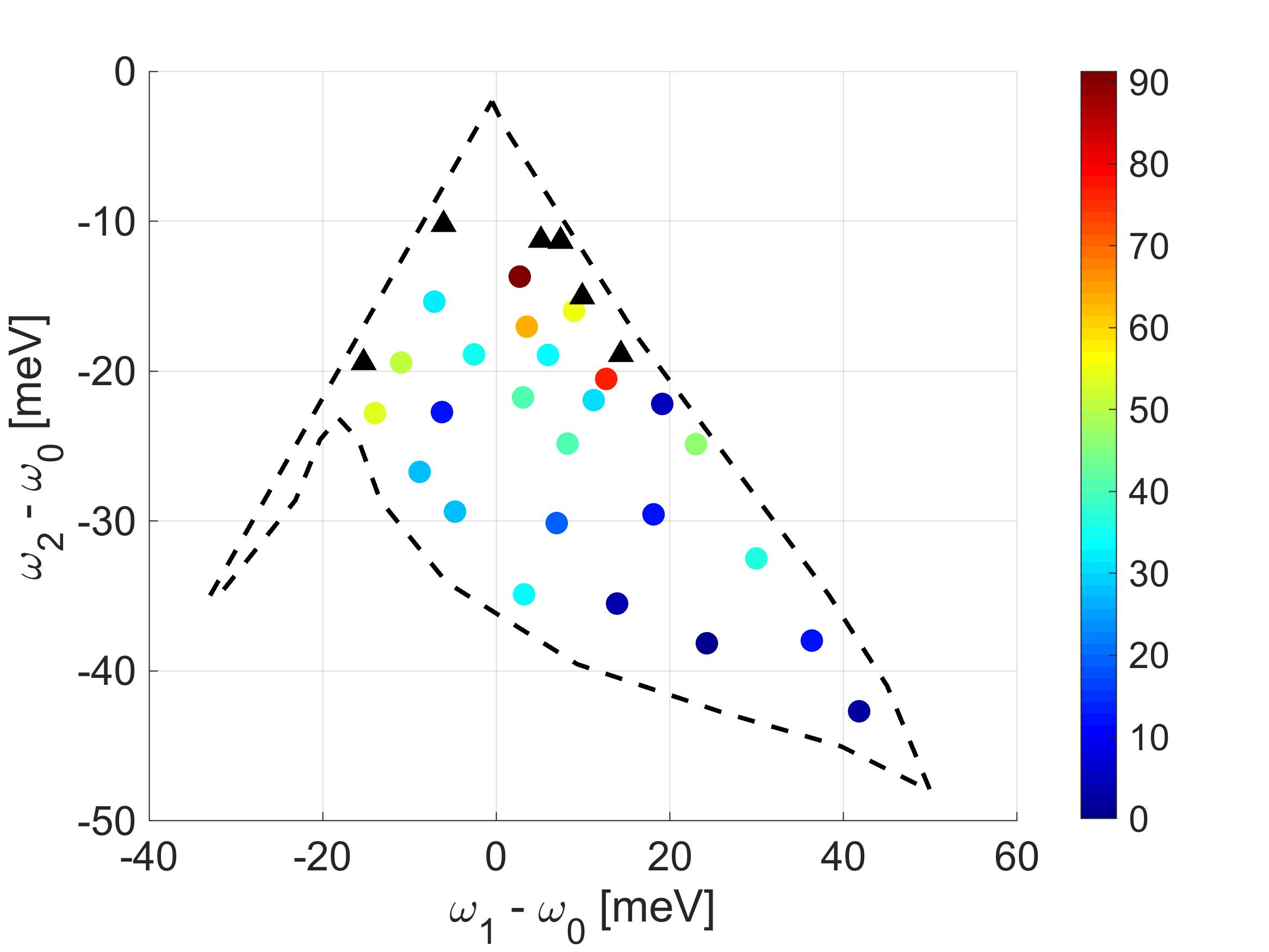}}
\caption{An overview of dual-wavelength solutions computed from initial conditions in Fig.~\ref{fig:surface}. The black dashed line is an outline of the lower part of Fig.~\ref{fig:surface}. Colored circles indicate stable dual-wavelength operation and black triangles indicate simulations that converge to single-wavelength operation. The color of the circles indicates the difference in relative spectral intensity between the two stable modes.}
\label{fig:surface2}
\end{figure}
The asymptotic dual-wavelength amplification is far from the linear background amplification, as seen in Fig.~\ref{fig:surface}. There is a broad range of potential dual-wavelength solutions with frequency separation ranging from around \unit[1]{THz} to above \unit[20]{THz} with multiple different field intensities. These solutions all distort the QW carriers to varying degrees depending on the field intensities and cavity field spectral locations. 

Generally, the successful simulations have a relatively short equilibration phase after initialization followed by a convergence into stable dual-wavelength operation. It is during this phase that the self-consistent nonlinear solution is generated from our \emph{ideal} initial guess in Fig.~\ref{fig:surface}. The non-equilibrium QW dynamics will to some degree pull the initial spectra away from the cold cavity modes, nonlinear wave mixing creates spectral peaks that feed back into the cavity, and the non-equilibrium modifications of the inversion can modify or even destabilize the dual-wavelength operation. Altogether, these effects can change the relative spectral peak intensities from the ideal initial guess or even lead to destabilization of the dual-mode operation.

Using the short-cavity mode stabilization, we find multiple independent solutions on the surface of Fig.~\ref{fig:surface} with frequency spacings from about \unit[4]{THz} and close to \unit[20]{THz}. In this range, the calculations converge in a manner similar to Fig.~\ref{fig:solutions}, but with different transient dynamics and final peak intensities. For spectral peak separations below \unit[4]{THz}, the longitudinal modes are too closely spaced and additional spectral filtering is required to stabilize dual-wavelength operation. In this frequency range, a FIR filter has been inserted into a longer external cavity to stabilize dual-wavelength solutions, with results similar to Fig.~\ref{fig:extraSolutions2}.

In Fig.~\ref{fig:surface2}, we present a summary of our full simulation results. In particular, we find that some mode-pairs near the red line, seen in Fig.~\ref{fig:surface}, converge directly to single-wavelength operation. This is because of nonlinear influences from the QWs that pull the wavelengths away from the initial guess and into a region with only single-mode operation. Furthermore, we find that the individual spectral peaks are not necessarily equally strong. In particular, we note that the deviations from our initial guesses tend to increase as we approach the single-wavelength region. 
Nevertheless, the our initialization method turns out to be very useful for reducing the parameter space of dual-wavelength solutions and reduces simulation time for very time consuming calculations. 

\section{Conclusion}
In summary, we have systematically investigated dual-wavelength operation in VECSELs using a microscopic many-body model with carrier scattering computed on the level of second Born-Markov approximation coupled to Maxwell's equation. In order to study dual-wavelength operation, an injection mapping scheme is used to approximate the asymptotic dual-wavelength operation.  This approximation is used to identify ranges of stable dual-wavelength operation in phase space and we implement a few different methods to realize select dual-wavelength output. 

Our numerical results show that dual-wavelength operation in VECSELs produces strong non-equilibrium QW carrier dynamics. The carriers are driven away from quasi-Fermi equilibrium dynamics and carrier scattering produces pronounced distortions in microscopic QW inversions. Even under stable dual-mode operation conditions, the carriers are not stationary but oscillate at the beating frequency.

Frequency selection via an intracavity etalon results in strong anti-correlated output intensity noise. The etalon does not produce clean dual-wavelength operation, but establishes two dominant families of modes. These modes compete for energy from energetically nearby carriers, which in turn create anti-correlated intensity fluctuations. Finally, we show that a promising method to stabilize the dual-wavelength output can be realized using an ideal frequency filter.

\section*{Funding}
This material is based upon work supported by the Air Force Office of Scientific Research under award number FA9550-17-1-0246.


\bibliography{sample}

\begin{thebibliography}{10}
\newcommand{\enquote}[1]{``#1''}

\bibitem{KellerTropper06}
U.~Keller and A.~C. Tropper, \enquote{Passively modelocked surface-emitting
  semiconductor lasers,} {\protect\JournalTitle{Phys. Rep.}} \textbf{429},
  67--120 (2006).

\bibitem{tropper2012ultrafast}
A.~C. Tropper, A.~H. Quarterman, and K.~G. Wilcox, \enquote{Ultrafast
  vertical-external-cavity surface-emitting semiconductor lasers,}
  {\protect\JournalTitle{Advances in Semiconductor Lasers}} \textbf{86},
  269--300 (2012).

\bibitem{tilma2015recent}
B.~W. Tilma, M.~Mangold, C.~A. Zaugg, S.~M. Link, D.~Waldburger, A.~Klenner,
  A.~S. Mayer, E.~Gini, M.~Golling, and U.~Keller, \enquote{Recent advances in
  ultrafast semiconductor disk lasers,} {\protect\JournalTitle{Light: Science
  \& Applications}} \textbf{4}, e310 (2015).

\bibitem{rahimi2016recent}
A.~Rahimi-Iman, \enquote{Recent advances in vecsels,}
  {\protect\JournalTitle{Journal of Optics}} \textbf{18}, 093003 (2016).

\bibitem{ElectronLett12}
B.~Heinen, T.-L. Wang, M.~Sparenberg, A.~Weber, B.~Kunert, J.~Hader, S.~W.
  Koch, J.~V. Moloney, M.~Koch, and W.~Stolz, \enquote{\unit[106]{W}
  continuous-wave output power from vertical-external-cavity surface-emitting
  laser,} {\protect\JournalTitle{Electron. Lett.}} \textbf{48}, 516 (2012).

\bibitem{LaserPhotonRev12}
T.-L. Wang, B.~Heinen, J.~Hader, C.~Dineen, M.~Sparenberg, A.~Weber, B.~Kunert,
  S.~W. Koch, J.~V. Moloney, M.~Koch, and W.~Stolz, \enquote{Quantum design
  strategy pushes high-power vertical-external-cavity surface-emitting lasers
  beyond \unit[100]{W},} {\protect\JournalTitle{Laser Photon. Rev.}}
  \textbf{6}, L12--L14 (2012).

\bibitem{zhang201423}
F.~Zhang, B.~Heinen, M.~Wichmann, C.~M{\"o}ller, B.~Kunert, A.~Rahimi-Iman,
  W.~Stolz, and M.~Koch, \enquote{A 23-watt single-frequency
  vertical-external-cavity surface-emitting laser,}
  {\protect\JournalTitle{Optics express}} \textbf{22}, 12817--12822 (2014).

\bibitem{waldburger2016high}
D.~Waldburger, S.~M. Link, M.~Mangold, C.~G.~E. Alfieri, E.~Gini, M.~Golling,
  B.~W. Tilma, and U.~Keller, \enquote{High-power \unit[100]{fs} semiconductor
  disk lasers,} {\protect\JournalTitle{Optica}} \textbf{3}, 844--852 (2016).

\bibitem{klopp13}
P.~Klopp, U.~Griebner, M.~Zorn, and M.~Weyers, \enquote{Pulse repetition rate
  up to \unit[92]{GHz} or pulse duration shorter than \unit[110]{fs} from a
  mode-locked semiconductor disk laser,} {\protect\JournalTitle{Appl. Phys.
  Lett.}} \textbf{98}, 071103 (2011).

\bibitem{laurain2018modeling}
A.~Laurain, I.~Kilen, J.~Hader, A.~Ruiz~Perez, P.~Ludewig, W.~Stolz,
  S.~Addamane, G.~Balakrishnan, S.~W. Koch, and J.~V. Moloney,
  \enquote{Modeling and experimental realization of modelocked vecsel producing
  high power sub-100 fs pulses,} {\protect\JournalTitle{Applied Physics
  Letters}} \textbf{113}, 121113 (2018).

\bibitem{axner1987detection}
O.~Axner, M.~Lejon, I.~Magnusson, H.~Rubinsztein-Dunlop, and
  S.~Sj{\"o}str{\"o}m, \enquote{Detection of traces in semiconductor materials
  by two-color laser-enhanced ionization spectroscopy in flames,}
  {\protect\JournalTitle{Applied optics}} \textbf{26}, 3521--3525 (1987).

\bibitem{zhou1993terahertz}
J.~Zhou, N.~Park, J.~W. Dawson, K.~J. Vahala, M.~A. Newkirk, and B.~I. Miller,
  \enquote{Terahertz four-wave mixing spectroscopy for study of ultrafast
  dynamics in a semiconductor optical amplifier,}
  {\protect\JournalTitle{Applied physics letters}} \textbf{63}, 1179--1181
  (1993).

\bibitem{langbein1990variations}
J.~O. Langbein, R.~O. Burford, and L.~E. Slater, \enquote{Variations in fault
  slip and strain accumulation at parkfield, california: Initial results using
  two-color geodimeter measurements, 1984--1988,}
  {\protect\JournalTitle{Journal of Geophysical Research: Solid Earth}}
  \textbf{95}, 2533--2552 (1990).

\bibitem{koch2004coherent}
G.~J. Koch, B.~W. Barnes, M.~Petros, J.~Y. Beyon, F.~Amzajerdian, J.~Yu, R.~E.
  Davis, S.~Ismail, S.~Vay, M.~J. Kavaya \emph{et~al.}, \enquote{Coherent
  differential absorption lidar measurements of co 2,}
  {\protect\JournalTitle{Applied optics}} \textbf{43}, 5092--5099 (2004).

\bibitem{jiang1993parameter}
S.~Jiang and M.~Dagenais, \enquote{Parameter extraction in semiconductor lasers
  using nearly degenerate four-wave mixing measurements,} in \emph{Lasers and
  Electro-Optics Society Annual Meeting, 1993. LEOS'93 Conference Proceedings.
  IEEE,}  (IEEE, 1993), pp. 578--579.

\bibitem{liu1994four}
J.-M. Liu and T.~B. Simpson, \enquote{Four-wave mixing and optical modulation
  in a semiconductor laser,} {\protect\JournalTitle{IEEE journal of quantum
  electronics}} \textbf{30}, 957--965 (1994).

\bibitem{kleine2001continuous}
T.~Kleine-Ostmann, P.~Knobloch, M.~Koch, S.~Hoffmann, M.~Breede, M.~Hofmann,
  G.~Hein, K.~Pierz, M.~Sperling, and K.~Donhuijsen, \enquote{Continuous-wave
  thz imaging,} {\protect\JournalTitle{Electronics Letters}} \textbf{37},
  1461--1463 (2001).

\bibitem{hu1995imaging}
B.~B. Hu and M.~C. Nuss, \enquote{Imaging with terahertz waves,}
  {\protect\JournalTitle{Optics letters}} \textbf{20}, 1716--1718 (1995).

\bibitem{yamaguchi2016brain}
S.~Yamaguchi, Y.~Fukushi, O.~Kubota, T.~Itsuji, T.~Ouchi, and S.~Yamamoto,
  \enquote{Brain tumor imaging of rat fresh tissue using terahertz
  spectroscopy,} {\protect\JournalTitle{Scientific reports}} \textbf{6} (2016).

\bibitem{federici2005thz}
J.~F. Federici, B.~Schulkin, F.~Huang, D.~Gary, R.~Barat, F.~Oliveira, and
  D.~Zimdars, \enquote{Thz imaging and sensing for security
  applications—explosives, weapons and drugs,}
  {\protect\JournalTitle{Semiconductor Science and Technology}} \textbf{20},
  S266 (2005).

\bibitem{wang1995tunable}
C.-L. Wang and C.-L. Pan, \enquote{Tunable multiterahertz beat signal
  generation from a two-wavelength laser-diode array,}
  {\protect\JournalTitle{Optics letters}} \textbf{20}, 1292--1294 (1995).

\bibitem{brown1995photomixing}
E.~R. Brown, K.~A. McIntosh, K.~B. Nichols, and C.~L. Dennis,
  \enquote{Photomixing up to 3.8 thz in low-temperature-grown gaas,}
  {\protect\JournalTitle{Applied Physics Letters}} \textbf{66}, 285--287
  (1995).

\bibitem{scheller2016dual}
M.~Scheller, C.~W. Baker, S.~W. Koch, and J.~V. Moloney,
  \enquote{Dual-wavelength passively mode-locked semiconductor disk laser,}
  {\protect\JournalTitle{IEEE Photonics Technology Letters}} \textbf{28},
  1325--1327.

\bibitem{scheller2010room}
M.~Scheller, J.~M. Yarborough, J.~V. Moloney, M.~Fallahi, M.~Koch, and S.~W.
  Koch, \enquote{Room temperature continuous wave milliwatt terahertz source,}
  {\protect\JournalTitle{Optics express}} \textbf{18}, 27112--27117 (2010).

\bibitem{leinonen2007dual}
T.~Leinonen, S.~Ranta, A.~Laakso, Y.~Morozov, M.~Saarinen, and M.~Pessa,
  \enquote{Dual-wavelength generation by vertical external cavity
  surface-emitting laser,} {\protect\JournalTitle{Optics express}} \textbf{15},
  13451--13456 (2007).

\bibitem{pal2010measurement}
V.~Pal, P.~Trofimoff, B.-X. Miranda, G.~Baili, M.~Alouini, L.~Morvan, D.~Dolfi,
  F.~Goldfarb, I.~Sagnes, R.~Ghosh \emph{et~al.}, \enquote{Measurement of the
  coupling constant in a two-frequency vecsel,} {\protect\JournalTitle{Optics
  express}} \textbf{18}, 5008--5014 (2010).

\bibitem{chernikov2012time}
A.~Chernikov, M.~Wichmann, M.~Shakfa, M.~Scheller, J.~Moloney, S.~Koch, and
  M.~Koch, \enquote{Time-dynamics of the two-color emission from
  vertical-external-cavity surface-emitting lasers,}
  {\protect\JournalTitle{Applied Physics Letters}} \textbf{100}, 041114 (2012).

\bibitem{wichmann2013evolution}
M.~Wichmann, M.~K. Shakfa, F.~Zhang, B.~Heinen, M.~Scheller, A.~Rahimi-Iman,
  W.~Stolz, J.~V. Moloney, S.~W. Koch, and M.~Koch, \enquote{Evolution of
  multi-mode operation in vertical-external-cavity surface-emitting lasers,}
  {\protect\JournalTitle{Optics Express}} \textbf{21}, 31940--31950 (2013).

\bibitem{wichmann2015antiphase}
M.~Wichmann, G.~Town, J.~Quante, M.~Gaafar, A.~Rahimi-Iman, W.~Stolz, S.~Koch,
  and M.~Koch, \enquote{Antiphase noise dynamics in a dual-wavelength
  vertical-external-cavity surface-emitting laser,} {\protect\JournalTitle{IEEE
  Photonics Technology Letters}} \textbf{27}, 2039--2042 (2015).

\bibitem{scheller2017high}
M.~Scheller, C.~W. Baker, S.~W. Koch, J.~V. Moloney, and R.~J. Jones,
  \enquote{High power dual-wavelength vecsel based on a multiple folded
  cavity,} {\protect\JournalTitle{IEEE Photonics Technology Letters}}
  \textbf{29}, 790--793 (2017).

\bibitem{de2013intensity}
S.~De, V.~Pal, A.~El~Amili, G.~Pillet, G.~Baili, M.~Alouini, I.~Sagnes,
  R.~Ghosh, and F.~Bretenaker, \enquote{Intensity noise correlations in a
  two-frequency vecsel,} {\protect\JournalTitle{Optics express}} \textbf{21},
  2538--2550 (2013).

\bibitem{koryukin2007antiphase}
I.~Koryukin and V.~Povyshev, \enquote{Antiphase dynamics of a multimode quantum
  well semiconductor laser,} {\protect\JournalTitle{Laser physics}}
  \textbf{17}, 680--683 (2007).

\bibitem{ahmed2002influence}
M.~Ahmed and M.~Yamada, \enquote{Influence of instantaneous mode competition on
  the dynamics of semiconductor lasers,} {\protect\JournalTitle{IEEE Journal of
  Quantum Electronics}} \textbf{38}, 682--693 (2002).

\bibitem{yacomotti2004dynamics}
A.~M. Yacomotti, L.~Furfaro, X.~Hachair, F.~Pedaci, M.~Giudici, J.~Tredicce,
  J.~Javaloyes, S.~Balle, E.~A. Viktorov, and P.~Mandel, \enquote{Dynamics of
  multimode semiconductor lasers,} {\protect\JournalTitle{Physical Review A}}
  \textbf{69}, 053816 (2004).

\bibitem{matus2004dynamics}
M.~Matus, M.~Kolesik, J.~V. Moloney, M.~Hofmann, and S.~W. Koch,
  \enquote{Dynamics of two-color laser systems with spectrally filtered
  feedback,} {\protect\JournalTitle{JOSA B}} \textbf{21}, 1758--1771 (2004).

\bibitem{baumner2011non}
A.~B{\"a}umner, S.~W. Koch, and J.~V. Moloney, \enquote{Non-equilibrium
  analysis of the two-color operation in semiconductor quantum-well lasers,}
  {\protect\JournalTitle{physica status solidi (b)}} \textbf{248}, 843--846
  (2011).

\bibitem{haug09}
H.~Haug and S.~W. Koch, \emph{Quantum Theory of the Optical and Electronic
  Properties of Semiconductors} (World Scientific, Singapore, 2009), 5th ed.

\bibitem{hader2003microscopic}
J.~Hader, S.~W. Koch, and J.~V. Moloney, \enquote{Microscopic theory of gain
  and spontaneous emission in gainnas laser material,}
  {\protect\JournalTitle{Solid-State Electronics}} \textbf{47}, 513--521
  (2003).

\bibitem{kilen2017non}
I.~Kilen, S.~W. Koch, J.~Hader, and J.~V. Moloney, \enquote{Non-equilibrium
  ultrashort pulse generation strategies in vecsels,}
  {\protect\JournalTitle{Optica}} \textbf{4}, 412--417 (2017).

\bibitem{kilen2018vecsel}
I.~Kilen, S.~W. Koch, J.~Hader, and J.~V. Moloney, \enquote{Vecsel design for
  high peak power ultrashort mode-locked operation,}
  {\protect\JournalTitle{Applied Physics Letters}} \textbf{112}, 262105 (2018).

\bibitem{kilen2014ultrafast}
I.~Kilen, J.~Hader, J.~V. Moloney, and S.~W. Koch, \enquote{Ultrafast
  nonequilibrium carrier dynamics in semiconductor laser mode locking,}
  {\protect\JournalTitle{Optica}} \textbf{1}, 192--197 (2014).

\bibitem{kilen2016fully}
I.~Kilen, S.~W. Koch, J.~Hader, and J.~V. Moloney, \enquote{Fully microscopic
  modeling of mode locking in microcavity lasers,} {\protect\JournalTitle{JOSA
  B}} \textbf{33}, 75--80 (2016).

\bibitem{kuhn1998density}
T.~Kuhn, \enquote{Density matrix theory of coherent ultrafast dynamics,} in
  \emph{Theory of transport properties of semiconductor nanostructures,}
  (Springer, 1998), pp. 173--214.

\bibitem{waldmuller2005intersubband}
I.~Waldm{\"u}ller, \emph{Intersubband Dynamics in Semiconductor Quantum
  Wells-Linear and Nonlinear Response of Quantum Confined Electrons}
  (Technische Universit{\"a}t Berlin, Berlin, 2005).

\bibitem{landolt1987numerical}
M.~Landolt and J.~B{\"o}rnstein, \enquote{Numerical data and functional
  relationships in science and technology, vol. 22/a of new series,}
  {\protect\JournalTitle{Group III}}  (1987).

\bibitem{hader2016ultrafast}
J.~Hader, M.~Scheller, A.~Laurain, I.~Kilen, C.~Baker, J.~Moloney, and S.~Koch,
  \enquote{Ultrafast non-equilibrium carrier dynamics in semiconductor laser
  mode-locking,} {\protect\JournalTitle{Semiconductor Science and Technology}}
  \textbf{32}, 013002 (2016).

\bibitem{smith1972surface}
J.~L. Smith, \enquote{Surface damage of gaas from 0.694-and 1.06-$\mu$ laser
  radiation,} {\protect\JournalTitle{Journal of Applied Physics}} \textbf{43},
  3399--3402 (1972).

\bibitem{sam1973laser}
C.~Sam, \enquote{Laser damage of gaas and znte at 1.06 $\mu$,}
  {\protect\JournalTitle{Applied optics}} \textbf{12}, 878--879 (1973).

\bibitem{qi2008investigation}
H.~Qi, Q.~Wang, X.~Zhang, Z.~Liu, Z.~Liu, J.~Chang, W.~Xia, and G.~Jin,
  \enquote{Investigation on damage process of gaas induced by 1064 nm
  continuous laser,} {\protect\JournalTitle{Journal of Applied Physics}}
  \textbf{103}, 033106 (2008).

\bibitem{breede2002fourier}
M.~Breede, S.~Hoffmann, J.~Zimmermann, J.~Struckmeier, M.~Hofmann,
  T.~Kleine-Ostmann, P.~Knobloch, M.~Koch, J.~Meyn, M.~Matus \emph{et~al.},
  \enquote{Fourier-transform external cavity lasers,}
  {\protect\JournalTitle{Optics communications}} \textbf{207}, 261--271 (2002).

\bibitem{wang1994tunable}
C.-L. Wang and C.-L. Pan, \enquote{Tunable dual-wavelength operation of a diode
  array with an external grating-loaded cavity,} {\protect\JournalTitle{Applied
  physics letters}} \textbf{64}, 3089--3091 (1994).

\bibitem{lee1997stable}
K.-S. Lee and C.~Shu, \enquote{Stable and widely tunable dual-wavelength
  continuous-wave operation of a semiconductor laser in a novel fabry-perot
  grating-lens external cavity,} {\protect\JournalTitle{IEEE journal of quantum
  electronics}} \textbf{33}, 1832--1838 (1997).

\end{thebibliography}

\end{document}